\documentclass[10pt,a4paper]{article}

\usepackage{url}
\usepackage{amsmath}
\usepackage{amsthm}
\usepackage{amsfonts}
\usepackage{graphicx}
\usepackage{float}
\usepackage{fancyhdr}
\usepackage[numbers]{natbib}
\newcommand{\bQ}{\mathbf{Q}}
\newcommand{\E}{\mathbf{E}}
\newcommand{\dd}{\mathrm{d}}
\date{07.07.2020}
\title{Volatility model calibration with neural networks \\ a comparison between direct and indirect methods}
\author{	
	 Dirk Roeder \footnote{The views expressed in this paper are solely the responsibility of the authors and should not be interpreted as reflecting the official positions of DZBANK. } \\ {\small dirk.roeder@dzbank.de}
	\and
	Georgi Dimitroff \footnote{The views expressed in this paper are solely the responsibility of the authors and should not be interpreted as reflecting the official positions of allianz global investors . } \\ {\small georgi.dimitroff@allianzgi.com}
}

\begin{document}

\maketitle

\medskip
\medskip

\section*{Abstract}

In a recent paper \cite{horvath2019deep} a fast 2-step deep calibration 
algorithm for rough volatility models was proposed: in the first step the time consuming mapping 
from the model parameter to the implied volatilities is learned by a  neural network 
and in the second step standard solver techniques are used to find the  best model parameter.

In our paper we compare these results with an alternative direct approach where the   the mapping from market implied volatilities to model parameters is  approximated by the neural network, without the  need for an extra solver step.
Using a whitening procedure and a projection of the target 
parameter to $[0,1]$, in order to be able to use  a sigmoid type output function we found that the direct approach outperforms the two-step one  for the data sets and methods published in \cite{horvath2019deep}. 

For our implementation we use  the  open source tensorflow 2 library \cite{tensorflow2015-whitepaper}.
The paper should be understood as a technical comparison of neural
network techniques and not as an methodically new Ansatz.

\medskip
\medskip
\medskip
\medskip

\vfill

\section{Introduction}

Calibrating the parameter of a volatility model to the market can be very time consuming, especially 
if there is no analytic solution for pricing the calibration products 
(mostly  plain-vanilla options), e.g. for the Rough Bergomi model \cite{BaFriJi2016}. Therefore a growing field 
of research is to use neural networks as part of the calibration algorithm to speed up the calibration process.

In \cite{horvath2019deep} the authors proposed a two step algorithm based on neural networks. In 
the first step the neural network is trained to predict the implied  volatilities from the 
volatility model parameter. Once the network is trained, the pricing can be done very 
efficiently as it is just a forward pass through the network. In the second step a  standard solver,  like the  Levenberg-Marquart, is used 
to calibrate the model, that is  to find the
volatility model parameter which minimize the reconstruction error between
the target/market volatilities and the predicted volatilities by the trained model. 
They found that the reconstruction errors are within the Monte Carlo error of the underlying
volatility model and solving can be done fast.

In a previous technical note \citep{SSRN:DimitRoedFrie2019} we have proposed an alternative approach where the neural network approximates the implicit mapping from the market implied volatilities to the optimal model parameters. In this case there is no need to wrap the neural network into an additional numerical solver in order to get the optimal model parameters and hence is more practicable, especially in a portfolio simulation context where one needs to calibrate derivative pricing models on each time step and path of a Monte Carlo simulation. 
In the context of the Heston model we have shown that the direct neural network calibration produces very accurate Heston model parameters. 

In the following we  show that, for the five data sets used in \cite{horvath2019deep}, 
a direct calibration of the  volatility model parameters to the market implied  volatilities 
can be done with a neural network very accurately and without over-fitting. The
big advantage is that, once the network is trained offline, no further online solver step
is necessary to find the parameter of the volatility model. 

The data sets and notebooks for \cite{horvath2019deep} can be found in their
github. Our alternative ansatz can be found in \cite{GITHUB:RoedDimit2020}. Throughout
the paper our method is referred as Volatilities-2-Model and the results from
\cite{horvath2019deep} as Model-2-Volatilities. 
\section{Two Volatility Model  Calibration Approaches  using Neural Network}\label{2nn_app}
The  no-arbitrage derivative pricing theory states that the price of an European style derivative can be calculated as a discounted risk-neutral  expectation   of the pay-off function. The pricing measure $\bQ$ used to calculate the risk-neutral expectation is unknown and needs to be estimated from the available market data. Typically $\bQ$ is modeled as the weak solution of a parameterized stochastic differential equation (SDE). In the case of the Rough Bergomi model the underlying $S_t$ under the pricing measure $\bQ$ follows the SDE  \cite{BaFriJi2016}:
\begin{align*}
   & \dd S_t = \mu_t S_t \dd t + \sigma_t S_t \dd Z_t\\
    & \sigma_t = \exp(X_t)\\
    & \dd X_t = \nu \dd W^ H_t - \alpha (X_t-m)\dd t
\end{align*}
where $\mu_t$ is an appropriate drift, like the repo-rate associated with the underlying $S_t$, $Z_t$ is a standard Brownian motion and $X_t$ is a fractional  Ornstein-Uhlenbeck process, that is $X_t$ satisfies an Ornstein-Uhlenbeck SDE  with respect to a fractional Brownian motion  $W^H_t$.

The parameters of any market  model are determined so that the observed market prices of plain-vanilla options are closely replicated by the model. 
To be more specific let us introduce some notation: denote the model parameters by $\theta$, and  the distribution of the solution of the model SDE by $\bQ(\theta)$.  
The (plain-vanilla) pricing function of the model is denoted by 
 $F^{\textbf{PVM}}$:
\begin{align*}
  F^\textbf{PVM}(\theta : K,T) =  D_T\E^{\bQ(\theta)}(S_T-K)^+
\end{align*}
where $D_T$ denotes the appropriate discounting factor depending on the deal's collateralisation  and $S_T$ is the underlying stock price  integration dummy variable denoting the solution of the modeling  SDE.  
Given the observed market prices $\mathbf{p}^{\text{mkt}}=\{p_i^{\text{mkt}}: i=1,\dots,n\}$  of call options\footnote{Obviously the observed market data contain call and put options among others, but without loss of generality for the sake of conciseness of the presentation we restrict ourselves to only call options.} with strikes and maturities $\{(K_i,T_i): i=1,\dots,n)\}$  the calibration of the parameter to the observed market data amounts to minimizing the loss $L$ over the parameter $\theta$:
\begin{align}\label{mkt_loss}
  L(\theta: \mathbf{p}^{\text{mkt}}) = \sum_{i=1}^n l(F^\textbf{PVM}(\theta : K_i,T_i), p_i^{\text{mkt}})\,,
\end{align}
where $l$ is some distance function, for example the squared distance $l(x,y)=(x-y)^2$.  

\noindent The solution of \eqref{mkt_loss} $\hat \theta $ can be viewed as a function of the market data mapping to the domain where the model parameters live:
\begin{align}\label{theta_hat}
 \hat \theta(\mathbf{p}^{\text{mkt}}) : \mathbf{p}^{\text{mkt}} \mapsto \text{argmin}_\theta   L(\theta: \mathbf{p}^{\text{mkt}})=:\hat \theta(\mathbf{p}^{\text{mkt}})\,.
\end{align}

In this setup there are at least two approaches to leverage the function-approximation capabilities of deep neural networks:
\begin{enumerate}
    \item Use a deep neural network to approximate the pricing function $F^\textbf{PVM}(\theta : K,T) $
    \item Use a deep neural network to approximate the calibrated model parameters function  $\hat \theta(\mathbf{p}^{\text{mkt}})$ from \eqref{theta_hat}
\end{enumerate}

In some models, like the prominent Heston model the pricing function is known in at least quasi-explicit form and the model to market loss $L$ can be  computed very efficiently. Therefore it would not make very much sense to try to approximate it using neural networks. In contrast, for some models we do not have  closed form pricing functions so using neural networks to  efficiently replicate  the Monte Carlo calculation of the plain vanilla calibration products is indeed a very sensible approach.   In the case of the Rough Bergomi model there is no known closed form for the pricing function so in  \cite{horvath2019deep} the neural network approximation of the $F^\textbf{PVM}(\theta : K,T) $ is proposed.

An important disadvantage of the first approach is that an additional numerical optimization algorithm needs to be used in order to calculate to model parameters $\hat \theta$. In the context of portfolio simulation  one needs to perform this numerical optimisation on each time discretization step and each Monte Carlo path. Even if the neural network approximation of $F^\textbf{PVM}(\theta : K,T)$, and hence the loss $L$ is  efficient, the numerical optimization within the Monte Carlo simulation becomes a significant computational bottleneck. 
On the other hand using a neural network to directly approximate the function 
\[\mathbf{p}^{\text{mkt}} \mapsto \hat \theta(\mathbf{p}^{\text{mkt}})\]
doesn't have to be wrapped with   an additional optimization algorithm. In \cite{SSRN:DimitRoedFrie2019}  the neural network approximation of  
$\hat \theta(\mathbf{p}^{\text{mkt}})$ in the case of the Heston stochastic volatility model was investigated and was  found to be of a very good quality. 

\noindent In this technical note we continue this line of work by  applying  the same direct approach in the context of the Rough Bergomi model  and compare  it against the two-step alternative  where one uses a neural network to approximate the pricing function $F^\textbf{PVM}(\theta : K,T)$.

\section{The Data sets and pre-processing}

To directly compare the both methods, we use the data sets and 
notebooks from \cite{horvath2019deep} as published in  github  and compare the results found for the 
train and test data sets with the results of our approach. 

There are five different data sets for 1. the Rough Bergomi model with flat forward variance, 
2. the Rough Bergomi model with picewise forward variance,
3. an one factor model with flat forward variance, 4. an one factor model with picewise forward variance, and
5. a Heston model. The summary of the number of samples and volatility model parameter per data set are 
summarized in table \ref{tab:datasets}.

\begin{table}[H]
\begin{center}
\begin{tabular}{ | l | l | l | l |}
\hline
 \textbf{data set name} & \textbf{train}& \textbf{test}& \textbf{model parameter} \\
\hline
	RoughBergomiFlatForwardVariance & 34000 & 6000 & 4\\
\hline
	RoughBergomiPicewiseForwardVariance & 68000 & 12000 & 11\\
\hline
	1FactorFlatForwardVariance & 34000 & 6000 & 4\\
\hline
	1FactorPiecewiseForwardVariance & 68000 & 12000 & 11\\
\hline
	Heston & 10200 & 1800 & 5\\
\hline
\end{tabular}
\caption{The number of items in the data sets for the five volatility models.}
	\label{tab:datasets}
\end{center}
\end{table}

\subsection{Input Data Pre-processing}
\begin{figure}[H]
	{\centering 
	\includegraphics[width=\textwidth]{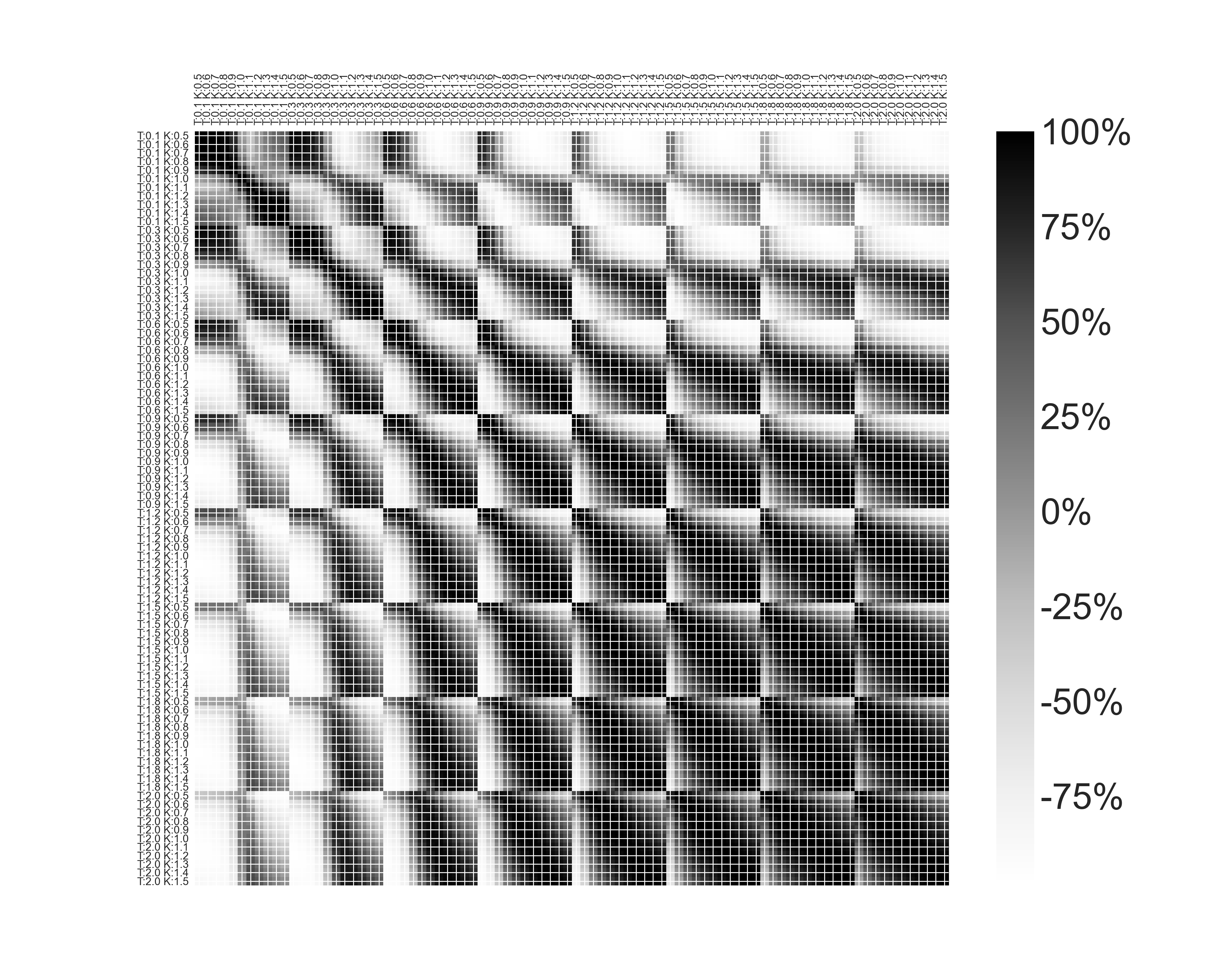}	
	\caption{The correlation matrix of the train data for RoughBergomiPicewiseForwardVariance 
	(with T the maturity and K the strike). }}
	\label{picture:RoughBergomiPiecewiseForwardVariance_COR}
	\end{figure}

	Before  constructing  the neural network, it is important  to have a closer look to  the input features
	and their inter-correlations. In figure \ref{picture:RoughBergomiPiecewiseForwardVariance_COR} the
	correlation matrix of the input features (the volatility surface) is shown for the Rough Bergomi model 
	with piecewise forward variance. It is not surprising that the implied volatilities on the surface are highly correlated and this correlation depends on the relative positions of the data points on the surface grid - obviously neighbouring volatilities express stronger correlations than the far off data points. For example  in the upper left corner the volatilities for the shortest maturity of 
	0.1 years and strike 0.6 is strongly  correlated with the one at strike  0.8 and more weakly correlated with  with the one at strike 1.4.
	
	The neural network is supposed to learn the correlations between the volatilities at different grid points of the volatility surface. One could support this process by adding  the 
	maturity and strike at each volatility instant into the input data, which is done e.g. in \cite{SSRN:DimitRoedFrie2019}. However  in the
	data sets used  here the strike-maturity   grid is fixed so we refrained from doing so. 
	
\begin{figure}[H]
	{\centering 	
	\includegraphics[width=\textwidth]{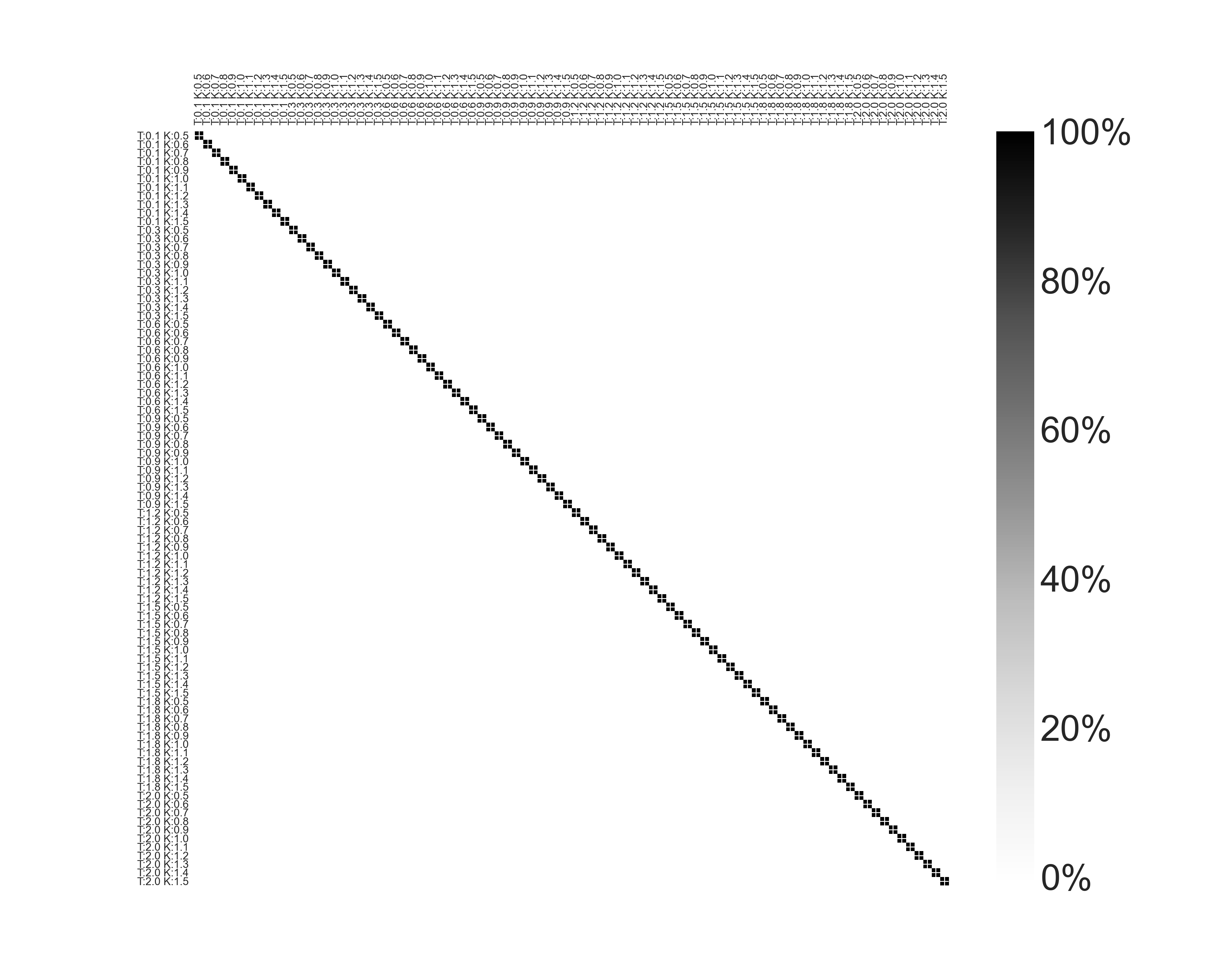}
	\caption{The correlation matrix of the train data for BergomiPicewiseForwardVariance after whitening 
	(with $T$ the maturity and $K$ the strike). }
	\label{picture:RoughBergomiPiecewiseForwardVariance_white_COR}
	}
	\end{figure}

	It is common also to standardize the data in order to numerically aid the training.
	We centered the data and instead of just  scaling, to get a unit sample variance, we used ZCA-Mahalanobis whitening in order to also de-correlate the input matrices. 
	In this process the centered input data are linearly transformed, that is multiplied by a de-correlation matrix $W$, so that  the sample correlation matrix  of the training data  is the identity. 
	For more on the  ZCA-Mahalanobis whitening procedure please refer to \cite{Kessy_2018}. We decided to use precisely this whitening approach because of its very natural property that the de-correlation is achieved by   a minimal additional adjustment, that is the  input data remain as close as possible in the $\mathrm{L}^2$-sense to the original input data (after centering of course).   
    The results of the correlation matrix after the whitening is shown 
	in figure \ref{picture:RoughBergomiPiecewiseForwardVariance_white_COR}.

\begin{figure}[H]
	{\centering 
	\includegraphics[width=0.45\textwidth]{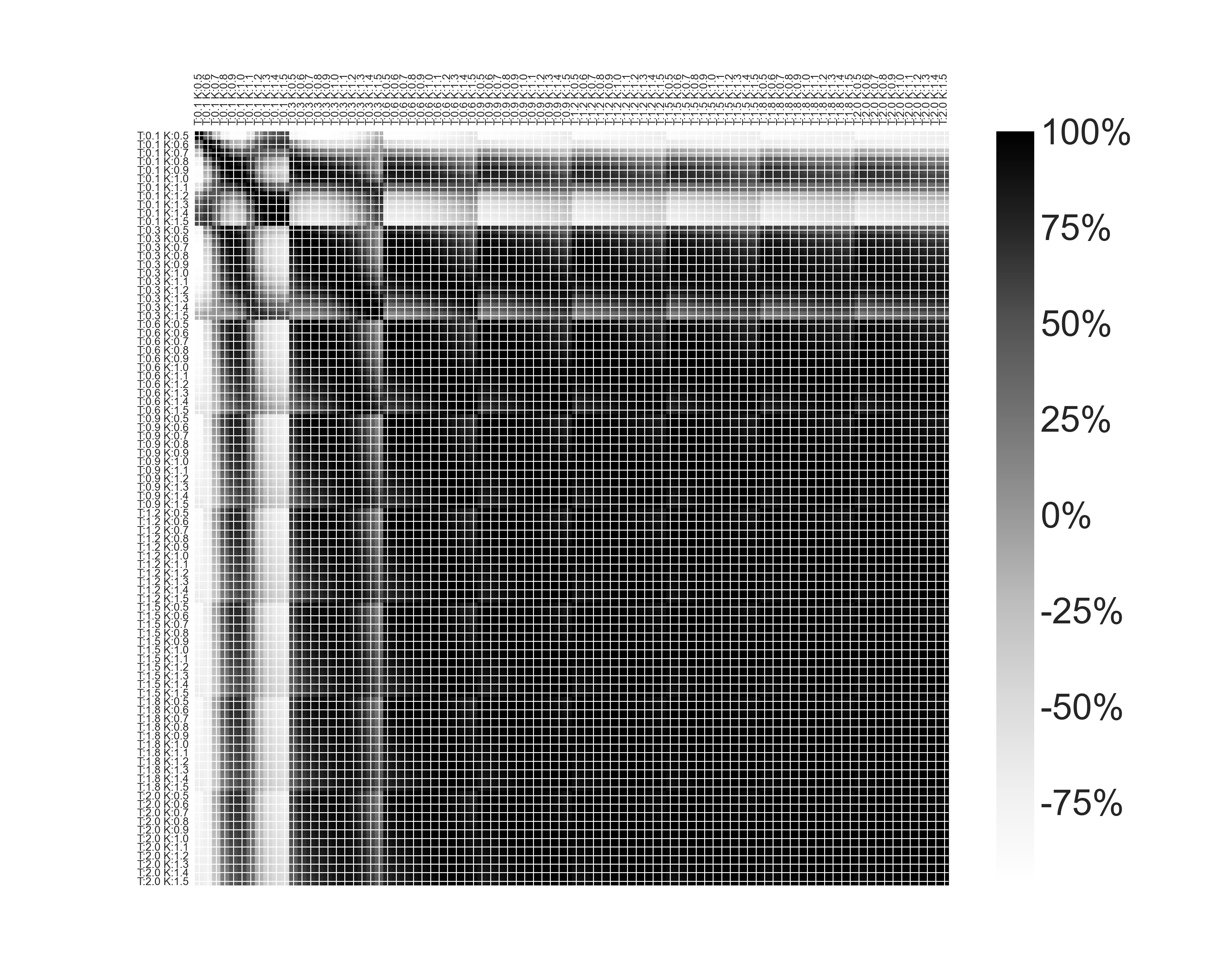}	
	\includegraphics[width=0.45\textwidth]{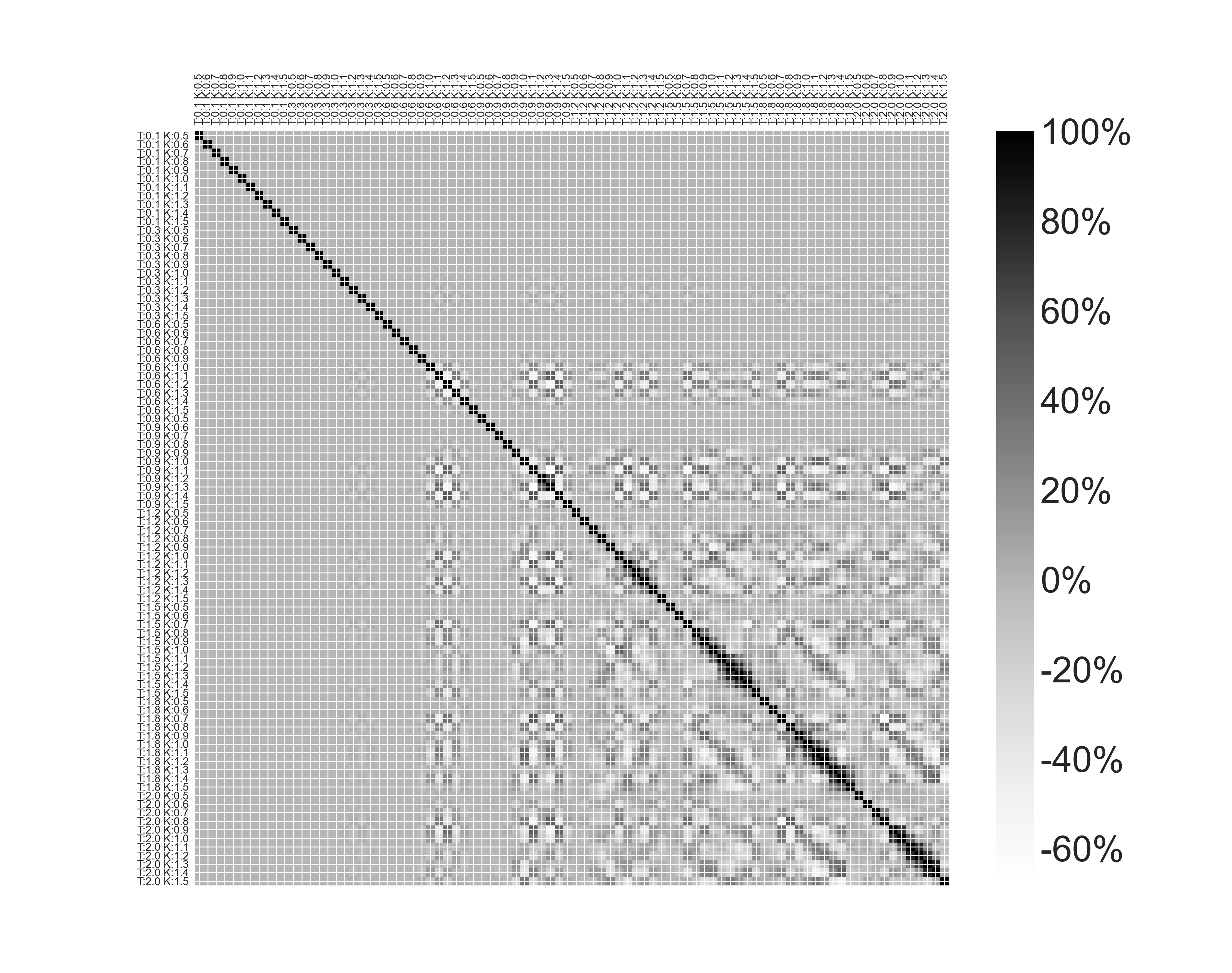}
	\caption{Left the correlation matrix of the train data for heston model before and right after whitening 
	(with T the maturity and K the strike). }
	\label{picture:Heston_COR}
	}
	\end{figure}

	The results after  whitening are very similar for the first four data sets. However for the 
	Heston data the correlation matrix is more problematic, and the whitening doesn't seem to work very well here, as can be seen in figure \ref{picture:Heston_COR}. Probably 
	this correlation structure is the reason that prediction results, shown later, are not 
	so good for the Heston data as for the other four models.
	
	Note that the matrix of the ZCA whitening, is constructed from the singular value decomposition of the sample covariance matrix of the training data which is then kept fixed and is being applied as such to the input data at inference time. 
	%the also a matrix constructed from the eigenvectors and a vector of the mean values)
	%are needed to predict the volatility parameter from input volatilities, so it becomes additional parameter of our method.

\section{The neural network architecture}

In the following chapter we explain how to construct the neural networks which are able to learn the implicit mapping from
the volatility surface directly to the parameter of the model. We will highlight the main points, all details 
can be found in the  jupyter notebooks made public  in the associated git repositories \cite{GITHUB:RoedDimit2020}.

\subsection{The neural network layer architecure}
\begin{figure}[H]
		{\centering 
		\includegraphics[width=0.5\textwidth]{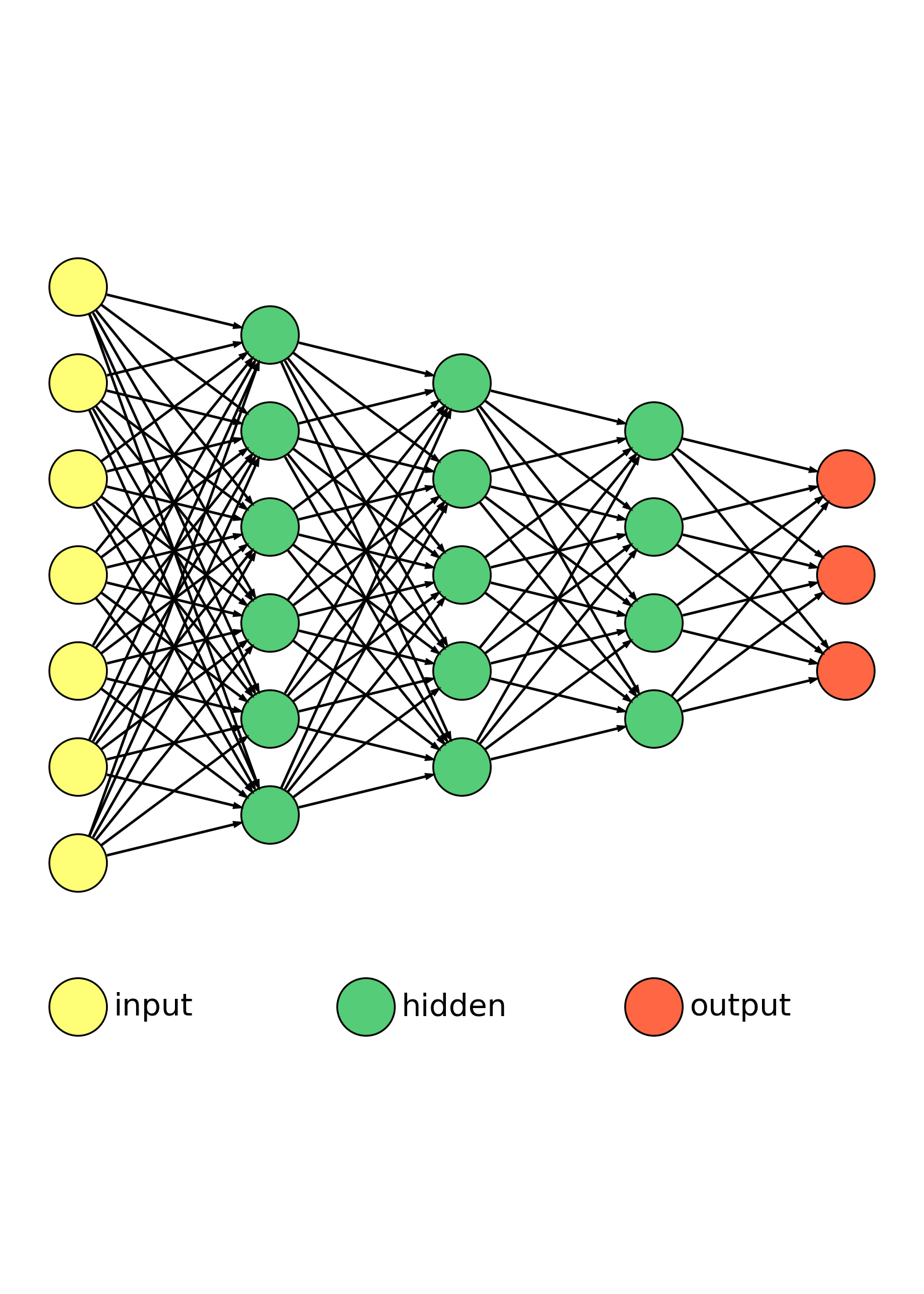}
		\caption{A schematic plot of the neural network architecture.}
		\label{picture:nnlayer}
		}
	\end{figure}

After pre-processing, in particular whitening, the input features are fed into a  simple feed forward network with fully connected layers as shown
	in figure \ref{picture:nnlayer}.  We found that three hidden layers with decreasing number 
	of neurons to be sufficient in order   to obtain excellent results. 
	For example for the Rough Bergomi model with piece-wise forward variance we use three hidden layers
	with 68, 49, and 30 neurons, which amount to 11274 calibration parameter of the neural network.
\subsection{The choice of the activation function}
\begin{figure}[H]
	{\centering 
	\includegraphics[width=\textwidth]{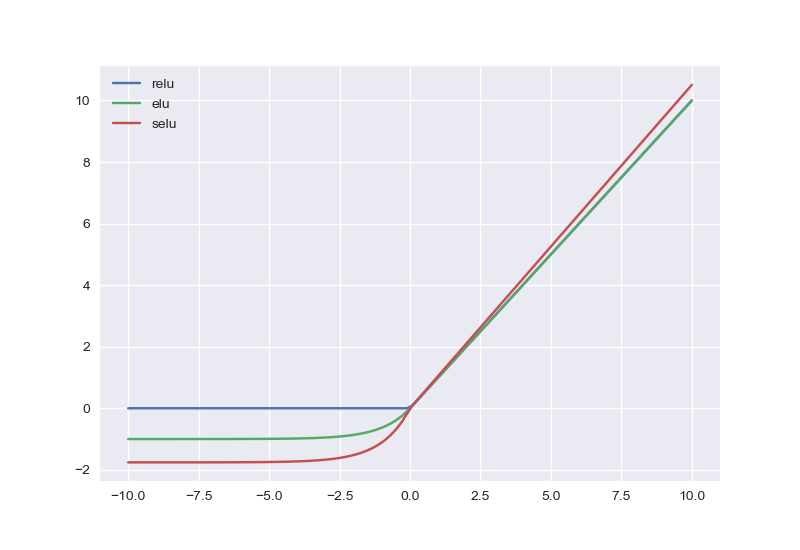}		
	\caption{Three popular activation functions.}
	\label{picture:reluEluSelu}
	}
	\end{figure}

	In figure \ref{picture:reluEluSelu} three popular activation functions are shown. For computer vision the \texttt{ReLU}  
	function is very widely  used because of its simplicity. 
	A slight modification	of this is the \texttt{eLU} function (used in \cite{horvath2019deep}) which has 
	the advantage that the negative values from the previous layer are not neglected. In our work we use another 
	modification,  the \texttt{SeLU}, a self normalized activation introduced in 2017 in \cite{DBLP:journals/corr/KlambauerUMH17}.
	The big advantage is that the \texttt{SeLU}-layers  tend to preserve the sample mean and variance of their respective inputs which leads to improved  training performance due to avoiding the vanishing gradient issues.  
\subsection{The output layer}
\begin{figure}[H]
	{\centering 
	\includegraphics[width=\textwidth]{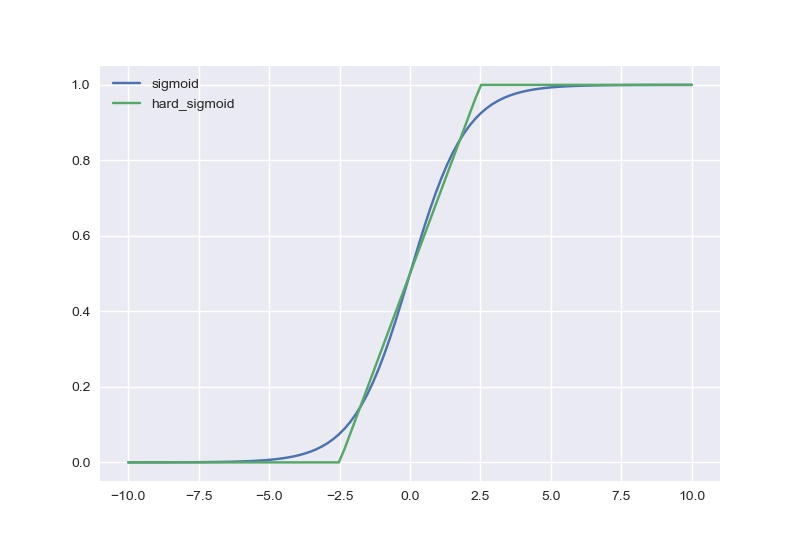}		
	\caption{The sigmoid functions.}
	\label{picture:sigmoid}
	}
	\end{figure}

The output of the neural network should lie into the parameter range expected by the volatility model. An easy way to force this is to  simply scale the 
target values (the parameter of the volatility model) to the unit interval  $[0,1]$ and respectively using a sigmoid output activation  function. Obviously, to obtain the volatility model parameters one needs to map  the predicted $[0,1]$ values back to the original parameter domain.  
For the scaling  we use $p_{[0,1]}=p\times(ub-lb)+lb$ with $ub$ the upper bound of the parameter $p$,
	$lb$ the lower bound and $p_{[0,1]}$ the transformed parameter. For numerical simplicity
	the hard sigmoid version, which are not smooth but numerical simpler and faster, can be
	used, cf. figure \ref{picture:sigmoid}. 
		
\subsection{Training }

	For the implementation and training of the neural network we use standard methods of tensorflow/keras. The adam optimizer as solver is  used where the training is performed on mini-batches. Early stopping was implemented in order to prevent over-fitting of the network.  
	
\begin{figure}[H]
	{\centering 
	\includegraphics[width=\textwidth]{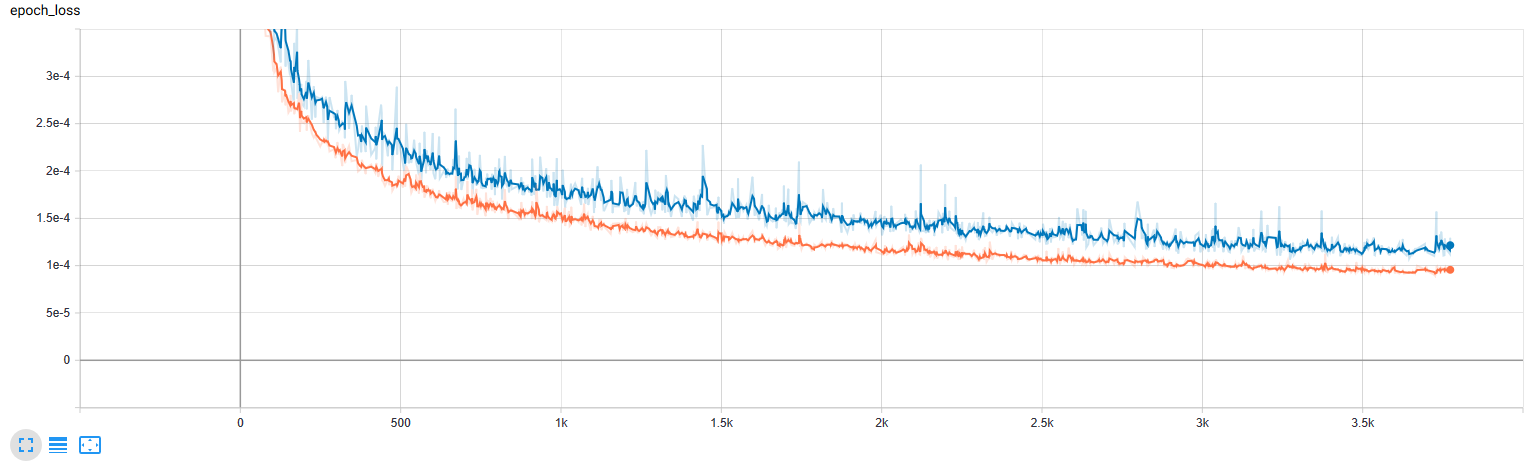}		
	\caption{The loss as a function of epochs for the training of the RoughBergomiPiecewiseForwardVariance data set. The
	blue line is the validation and the orange one the train set.}
	\label{picture:RoughBergomiPiecewiseForwardVariance_Loss}
	}
	\end{figure}

	As loss function we use the mean squared error between the target and predicted volatility model parameters, cf. fig. \ref{picture:RoughBergomiPiecewiseForwardVariance_Loss}.	
	Again, all technical details can be found on the Jupyter notebook in our github repository \cite{GITHUB:RoedDimit2020}.
			
\section{Results}

\begin{figure}[H]
{\centering 
\fbox{\includegraphics[width=0.45\textwidth]{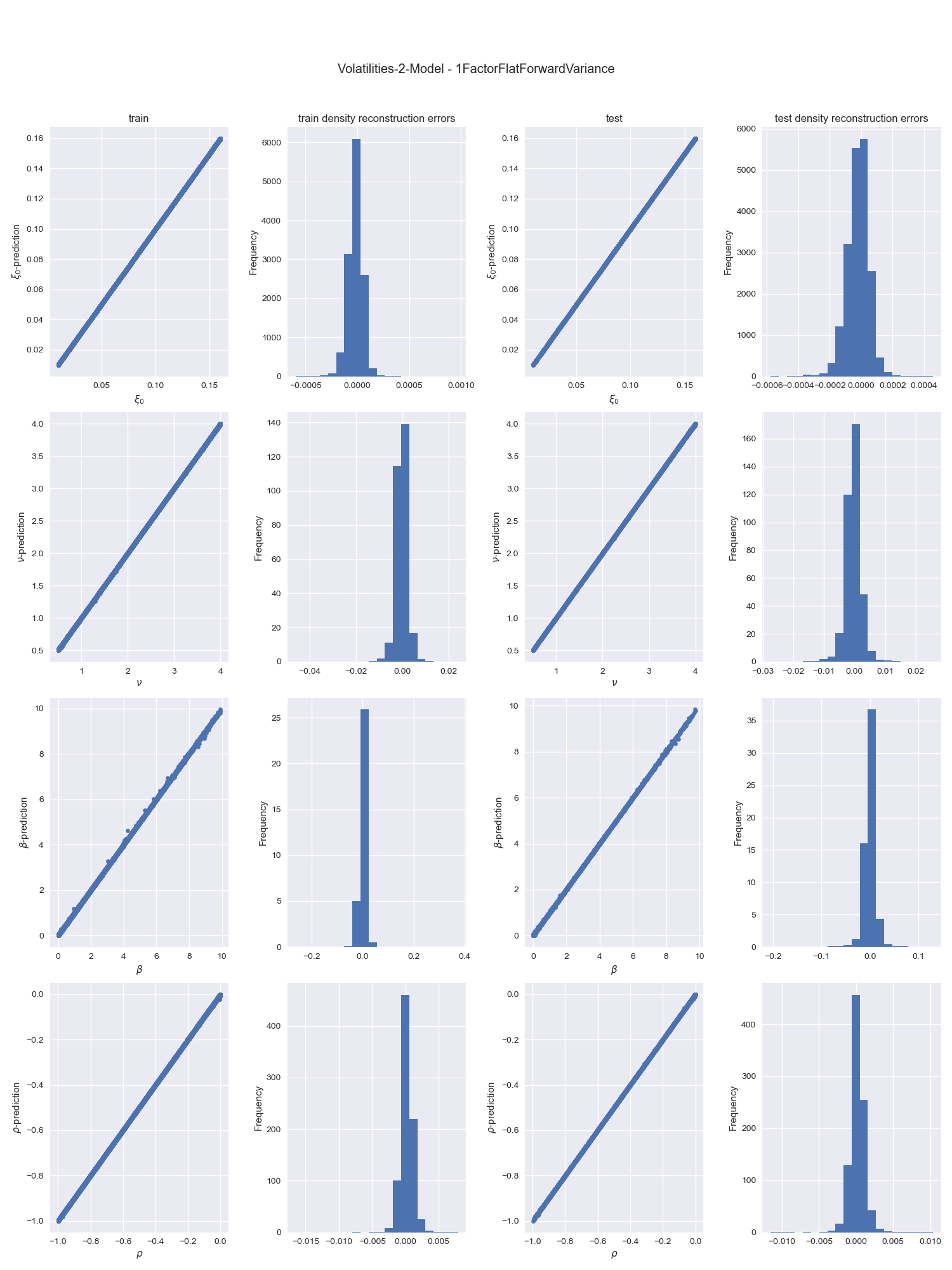}}
\fbox{\includegraphics[width=0.45\textwidth]{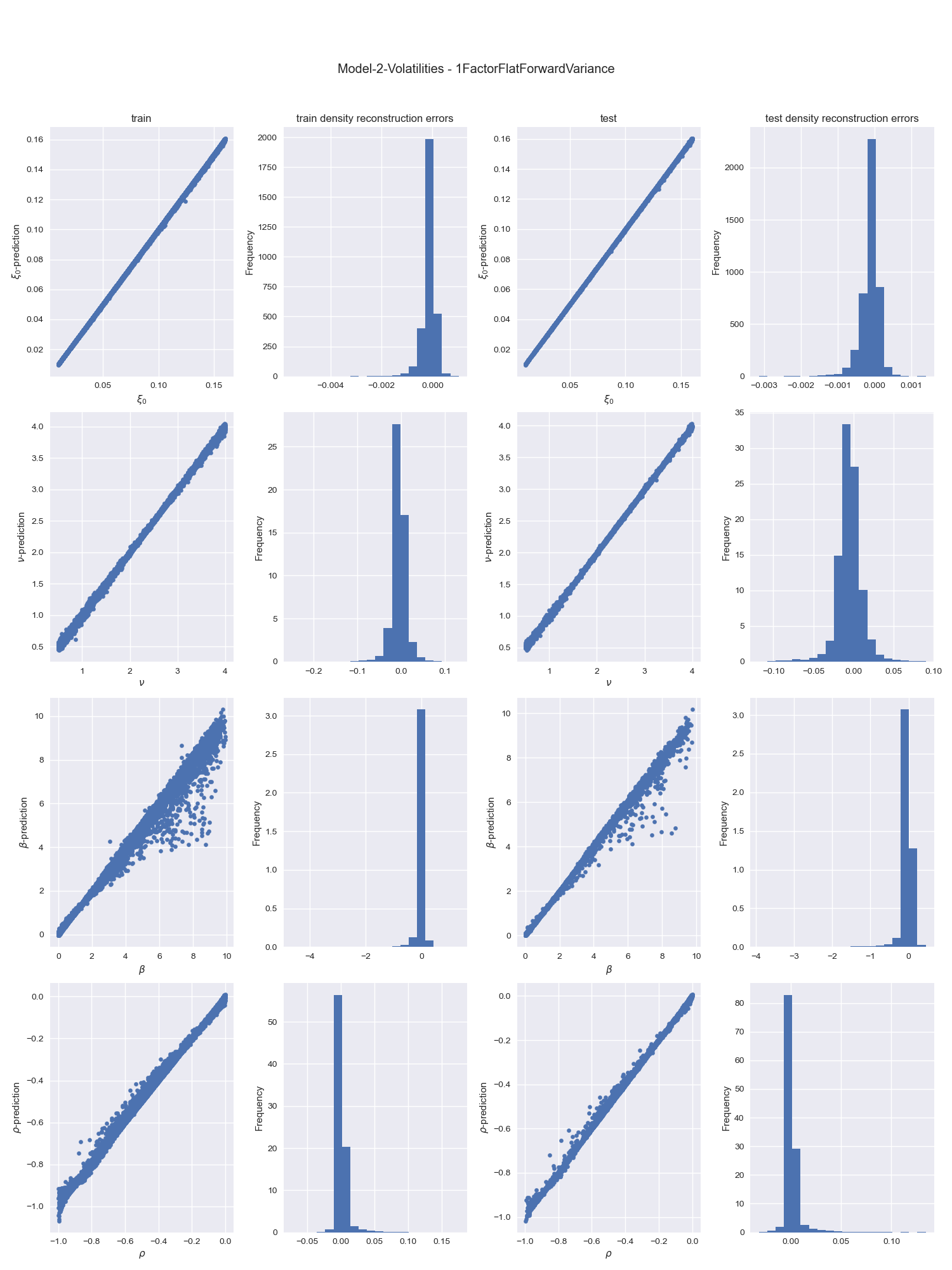}}
\caption{1FactorFlatForwardVariance}
\label{picture:1FactorFlatForwardVariance}
}
\end{figure}

\begin{figure}[H]
{\centering 
\fbox{\includegraphics[width=0.45\textwidth]{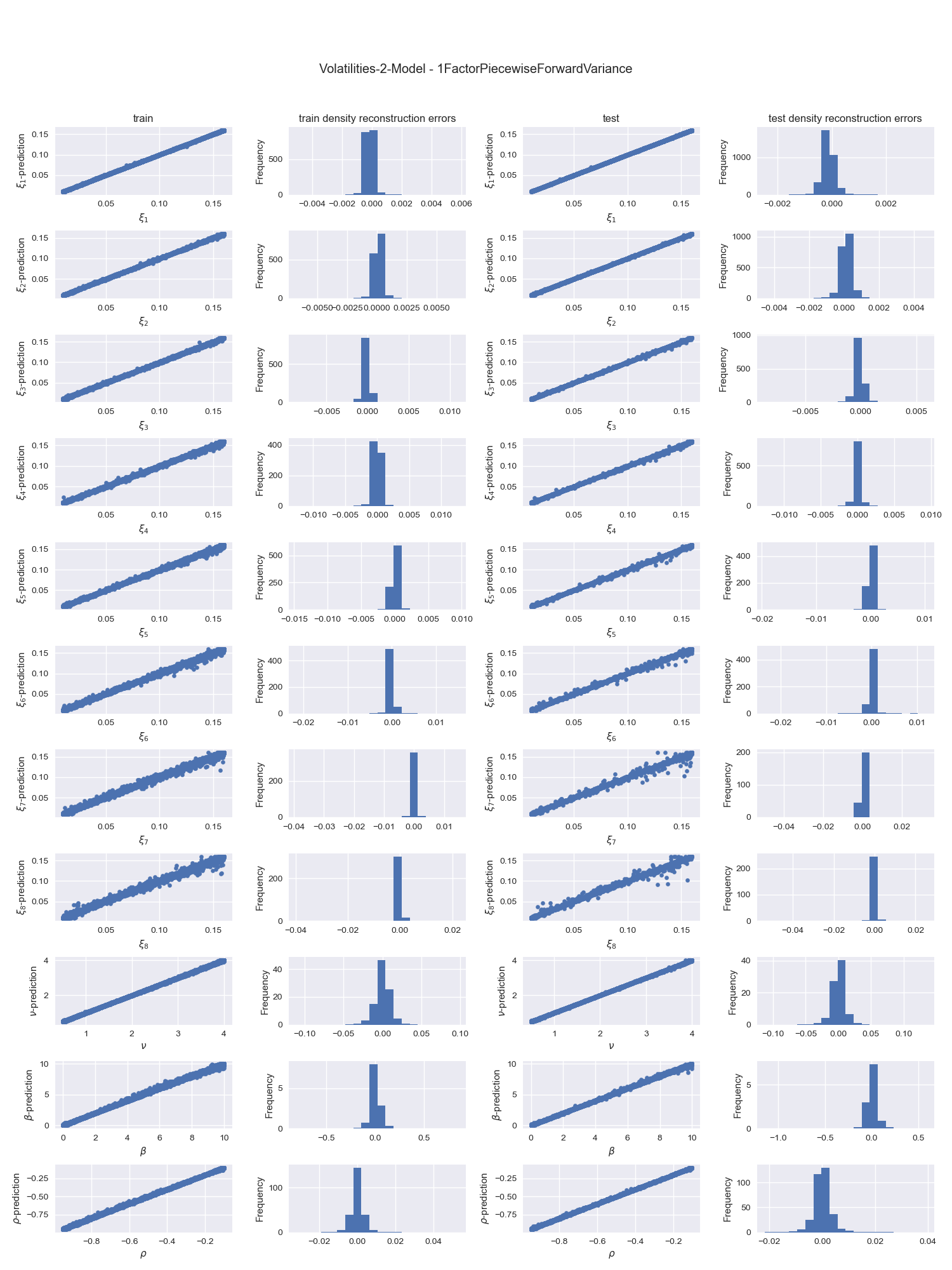}}
\fbox{\includegraphics[width=0.45\textwidth]{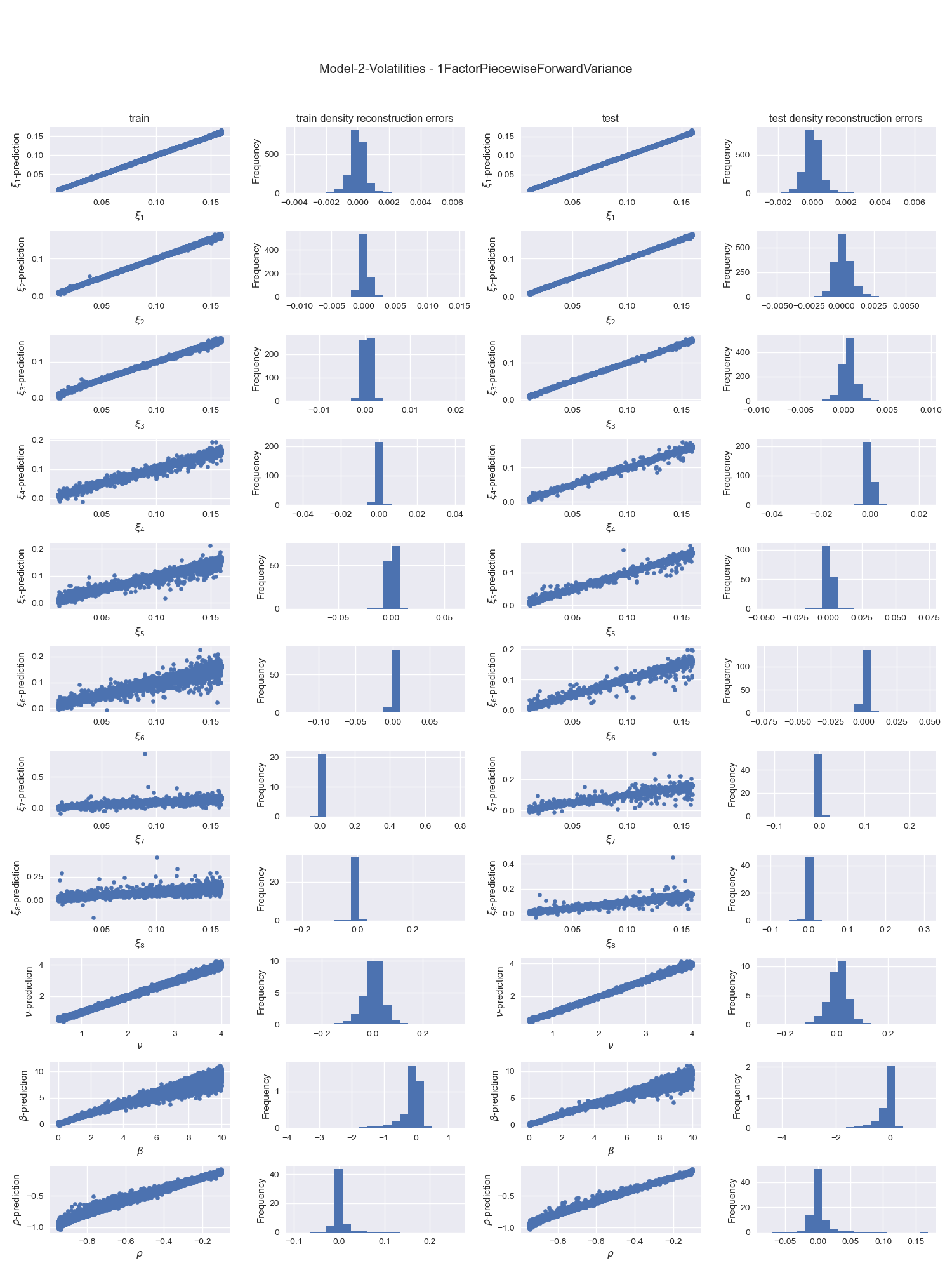}}
\caption{1FactorPiecewiseForwardVariance.}
\label{picture:1FactorPiecewiseForwardVariance}
}
\end{figure}

\begin{figure}[H]
{\centering 
\fbox{\includegraphics[width=0.45\textwidth]{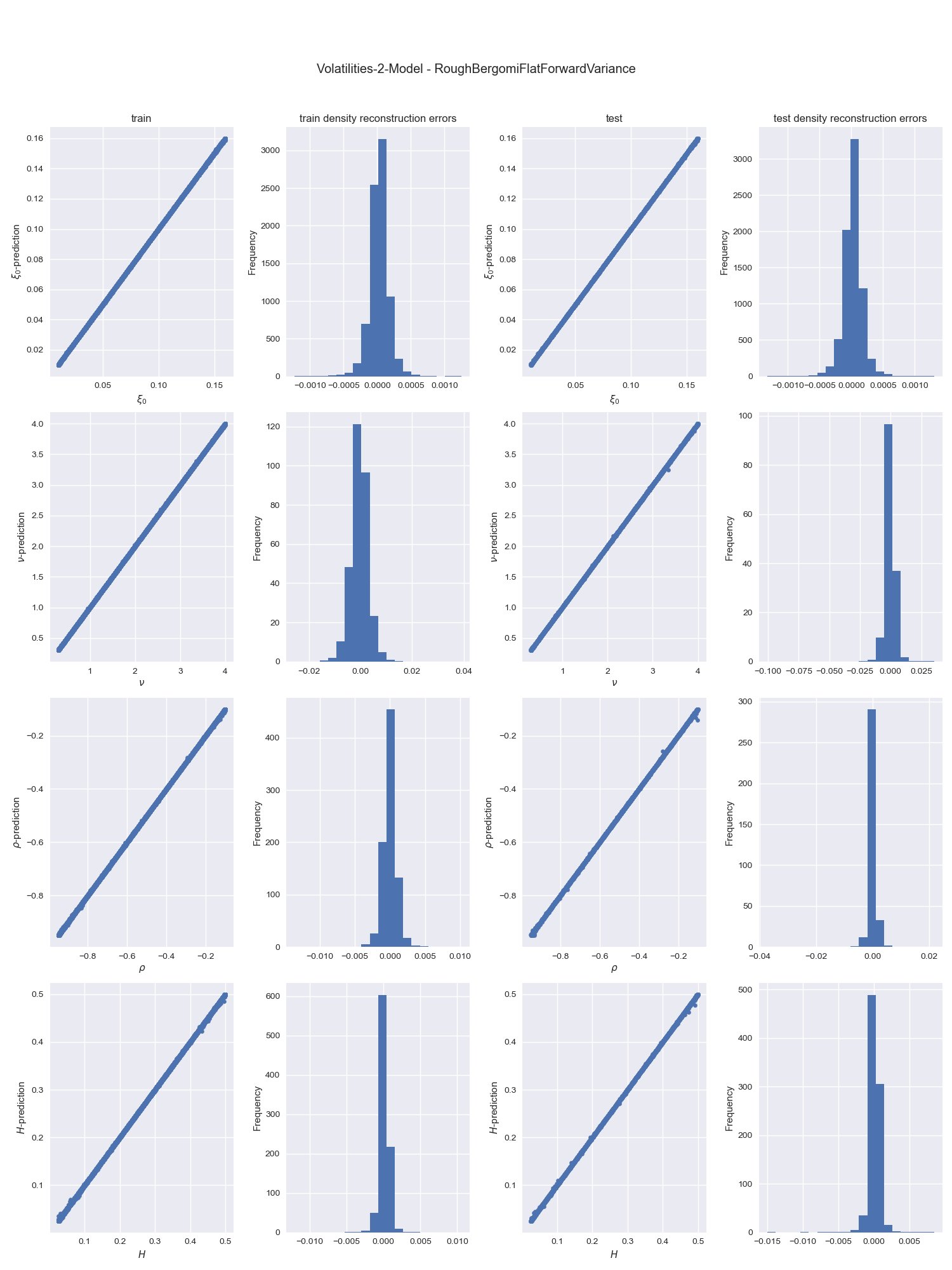}}
\fbox{\includegraphics[width=0.45\textwidth]{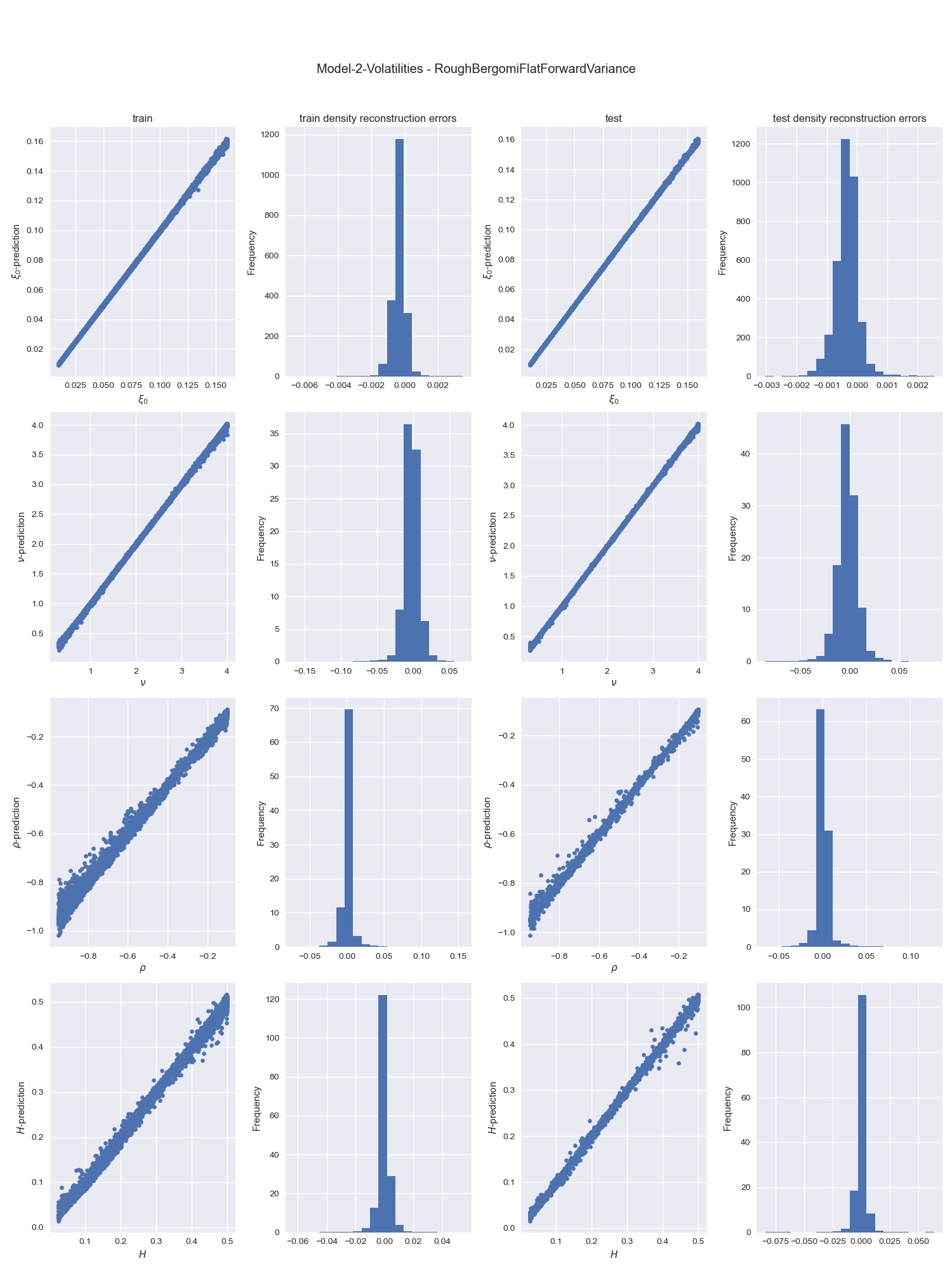}}
\caption{RoughBergomiFlatForwardVariance.}
\label{picture:RoughBergomiFlatForwardVariance}
}
\end{figure}

\begin{figure}[H]
{\centering 
\fbox{\includegraphics[width=0.45\textwidth]{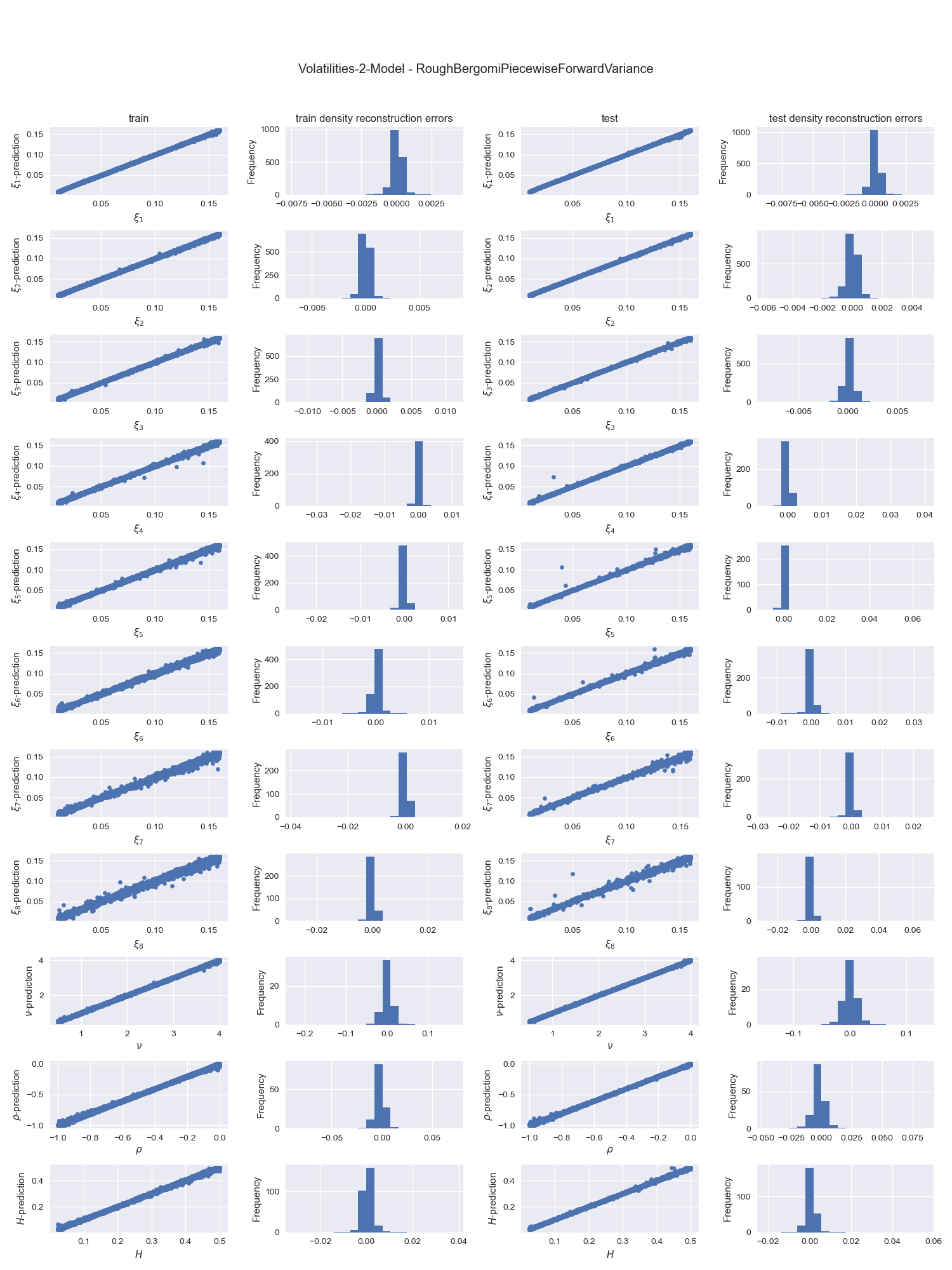}}
\fbox{\includegraphics[width=0.45\textwidth]{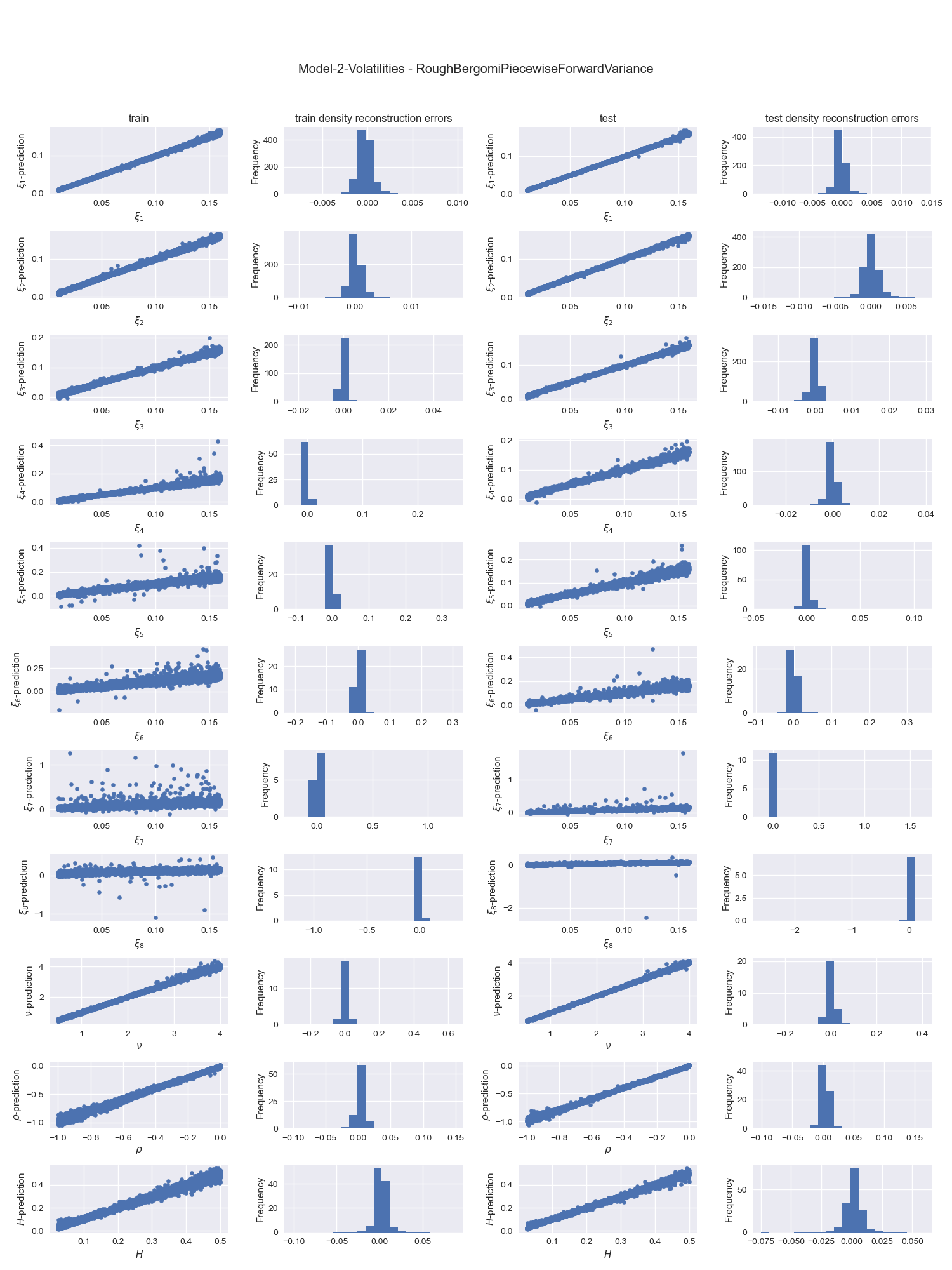}}
\caption{RoughBergomiPiecewiseForwardVariance.}
\label{picture:RoughBergomiPiecewiseForwardVariance}
}
\end{figure}

\begin{figure}[H]
{\centering 
\fbox{\includegraphics[width=0.45\textwidth]{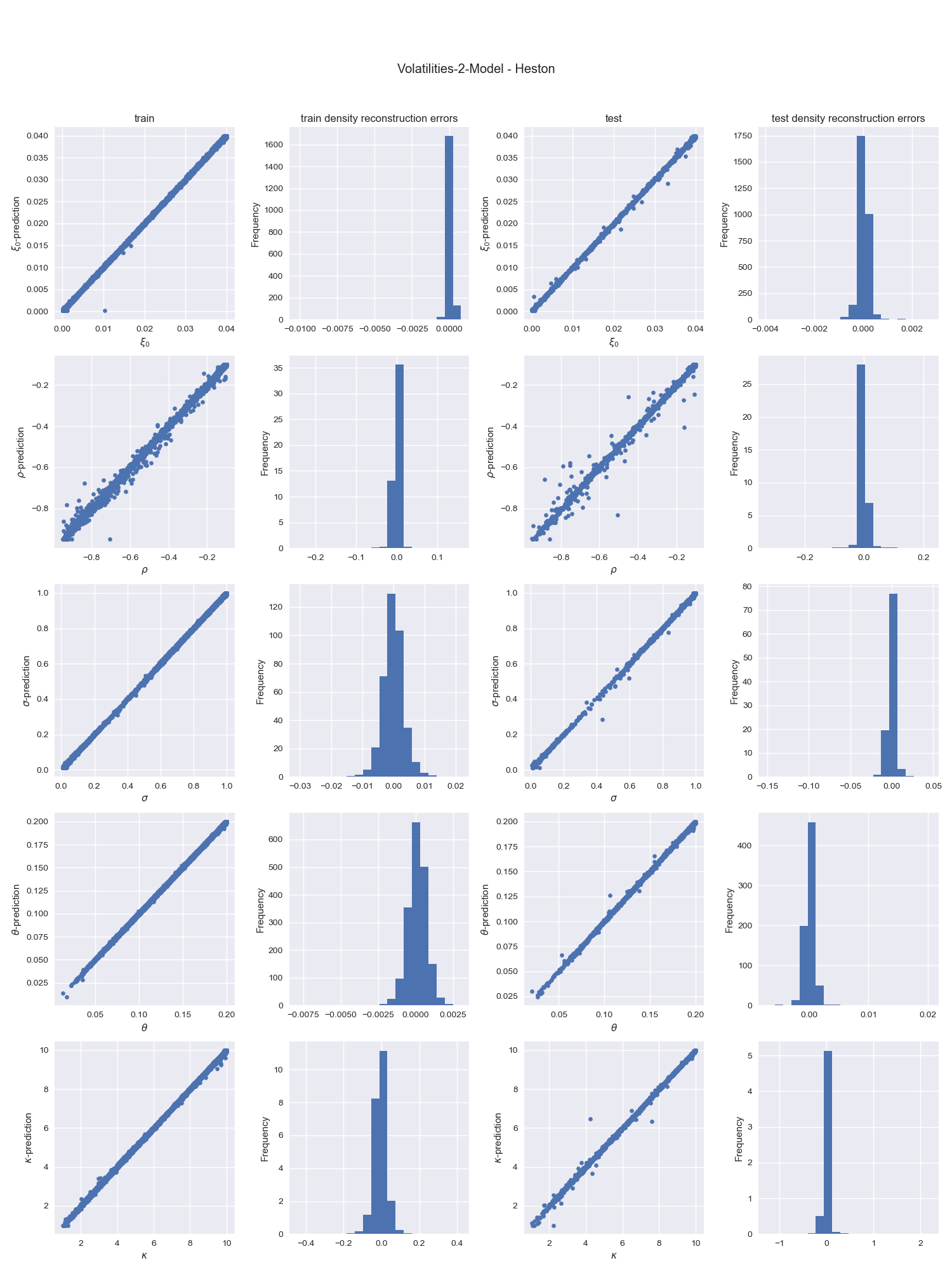}}
\fbox{\includegraphics[width=0.45\textwidth]{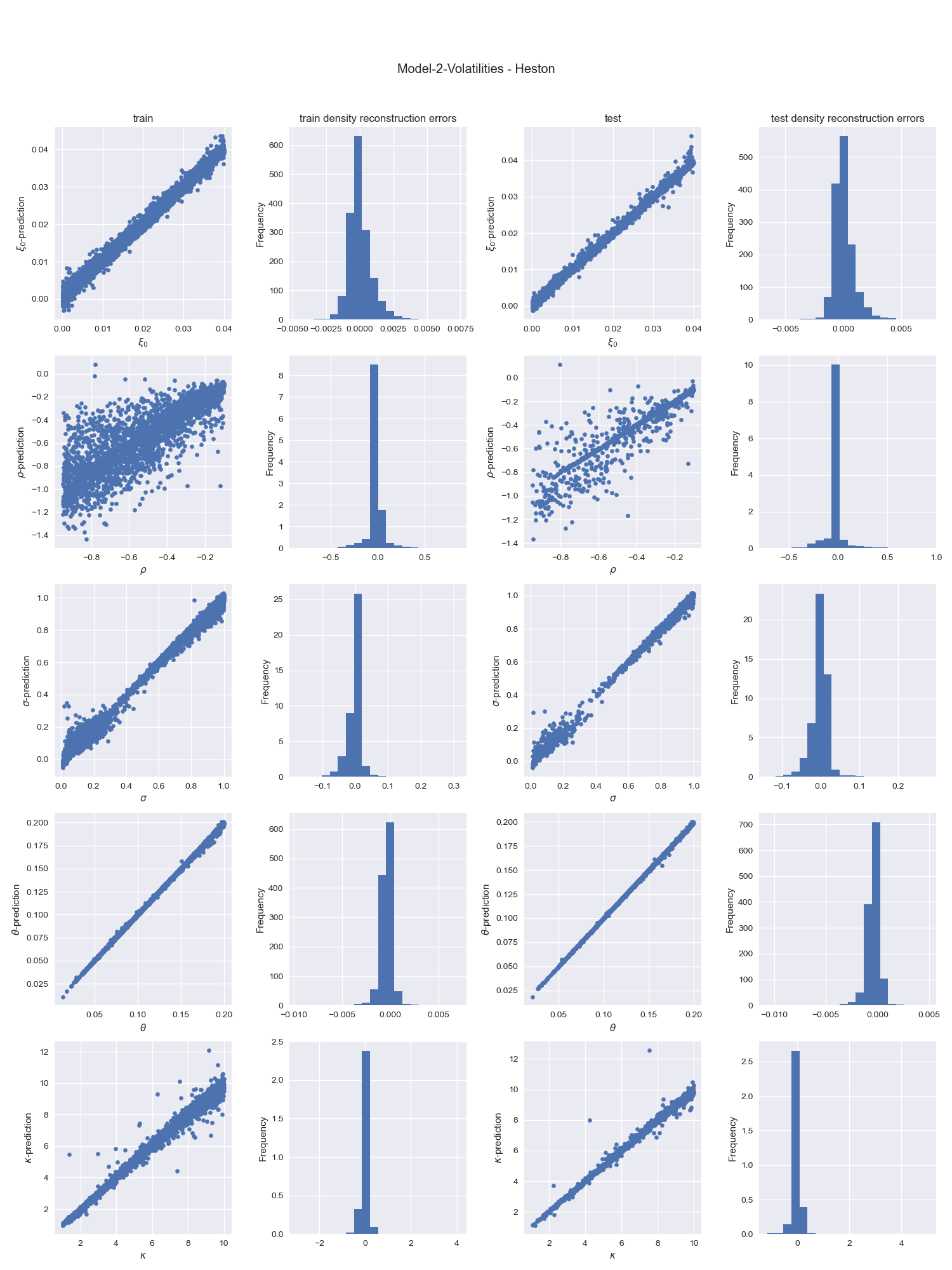}}
\caption{Heston.}
\label{picture:Heston}
}
\end{figure}

In figure \ref{picture:1FactorFlatForwardVariance}, \ref{picture:1FactorPiecewiseForwardVariance},
\ref{picture:RoughBergomiFlatForwardVariance}, \ref{picture:RoughBergomiPiecewiseForwardVariance}, and
\ref{picture:Heston} the results for the five datasets (cf. table \ref{tab:datasets})
and the two methods, Volatilities-2-Model (left part of the pictures) and Model-2-Volatilities
\footnote{In \cite{horvath2019deep} different solver are compared but the results are very similar, 
here the results for Levenberg Marquardt are shown.}
 (right part) are shown. 
The rows represents the different parameter of the volatility model, e.g. figure \ref{picture:1FactorFlatForwardVariance}
from top to button  $\xi_0$, $\nu$ ,$\beta$, and $\rho$. In the first column of the left (right) picture the
target vs. the predicted parameter are plotted for all training data sets with the Volatilities-2-Model (Model-2-Volatilities)
Ansatz. The second column shows the reconstruction loss from the the first 
column\footnote{The difference between the target and predicted
parameter} as a density (the integral is normalized to one). The third and fourth column shows the 
same for the test data.

As one can see the predictions are indeed very close to   the target parameters, where the quality of the direct implied vol to parameter approach, dubbed the Volatilities-2-Model Ansatz, is superior. On the test data we see the same performance as on the train subset, meaning that the network is able to generalize to unseen data, they do not experience over-fitting issues.

Remarkable is that the results for the Heston model in figure \ref{picture:Heston} tend to be slightly worse than for the Rough Bergomi model. As mentioned above, maybe this is an effect of the correlation structure of the volatilities used here as training data. Another issue with the Heston model can be the model parameter identifiability - there are  Heston parameterizations which differ substantially on the values of the Heston parameters but  correspond to extremely similar implied volatility surfaces.

\section{Conclusion and remarks}

There are two main approaches for aiding a volatility model calibration with deep  neural networks: 
\begin{enumerate}
    \item Use the neural network to directly learn the implicit mapping from the market implied vols to the volatility model parameters
    \item Use the network to learn the pricing function of the model, that is the function mapping the model parameters, strike and maturity  into the corresponding implied volatility. Then use this approximation within a numerical optimization routine, aka solver, in order to calibrate the model parameters given the market volatilities. 
\end{enumerate}

We have  implemented the first approach and compared its performance with the second approach followed in \cite{horvath2019deep}.   The 
predictions using the first, direct, approach are superior. Our experiments also show that the direct market vol to model parameter neural network generalizes well to unseen data.  Additionally, from the computational perspective the  first approach is also to be preferred since it does not require an external solver loop. 

We found that the whitening of the highly correlated vol surface inputs leads to a more fast and stable training.  The scaling of the  target parameter to the unit $[0,1]$
interval and using  a  sigmoid-like output activations   forced the 
predicted parameter to lie within the target boundaries of the model and hence improved the interpretability and usability of the results. 

Note that the parameter sets used here are generated with the models themselves,
that is  every volatility surface perfectly fits  to a valid model parameter set by construction. What
would happen if the target volatility model cannot fit the volatility surface
shown in real market?  In such cases it would be beneficial to bias the network towards fitting the more liquid sections of the volatility surface, as  the neural network has no information on what ranges of the surface are more important. In practical  situations
 ATM volatilities are more important than  far out of the money
ones. All this is routinely  done in the daily calibration process in  financial 
institutions  for example by adding more weight to ATM calibration. To train a neural network to 
reflect these requirements one way is to use the 
standard calibration process in order to generate training data. In \cite{SSRN:DimitRoedFrie2019} we used such data 
for a Heston model with very good results. 

Certainly one shouldn't trust the results without back-testing, i.e. if one 
would like to replace her  calibration procedure with a fast neural network
approach then the calibration error should be monitored on a regular basis.    

\bibliographystyle{unsrt}
\bibliography{bibliography}

\end{document}